\documentclass[aps,pra,twocolumn,showpacs,superscriptaddress,floatfix]{revtex4}

\usepackage{amsmath}
\usepackage{amssymb}
\usepackage{amsbsy}
\usepackage{euscript}

\usepackage{graphicx}
\usepackage[usenames]{color}

\newcommand{\be}{\begin{equation}}
\newcommand{\ee}{\end{equation}}

\begin{document}

\title{Cross sections for short pulse single and double ionization of helium}

\author{A. Palacios}
\affiliation{Lawrence Berkeley National Laboratory, Chemical Sciences, Berkeley, CA 94720}
\author{T. N. Rescigno}
\affiliation{Lawrence Berkeley National Laboratory, Chemical Sciences, Berkeley, CA 94720}
\author{C.W. McCurdy}
\affiliation{Lawrence Berkeley National Laboratory, Chemical Sciences, Berkeley, CA 94720}
\affiliation{Departments of Applied Science and Chemistry, University of California, Davis,  CA 95616}

\date{\today}
\begin{abstract}
In a previous publication, procedures were proposed for unambiguously extracting amplitudes for single and double ionization from a time-dependent wavepacket by effectively propagating for an infinite time following a radiation pulse.  Here we demonstrate the accuracy and utility of those methods for describing two-photon single and one-photon double ionization of helium.  In particular it is shown how narrow features corresponding to autoionizing states are easily resolved with these methods.
\end{abstract}

\maketitle

\section{Introduction}
The development of attosecond-pulse radiation sources ~\cite{agostini04} offers the prospect of a new class of pump/probe experiments that can in principle  explore the effects of electron correlation in atoms and molecules on  ultrashort time scales ~\cite{ergler06,niikura06,kienberger02}. The analysis and interpretation of such  experiments will necessarily involve state-of-the-art time-dependent, non-perturbative theoretical methods and advanced supercomputing resources. For example, if the pump step excites autoionizing states of the target whose lifetimes are much longer than the pulse duration, then traditional time-dependent methods might require prohibitively long integration times to compute meaningful ionization probabilities.  

In a previous paper~\cite{palacios07}, hereafter referred to as paper I, we outlined a procedure for extracting the amplitudes for ejecting electrons of particular energies and directions from a quantum wavepacket at the end of a short pulse, while the electrons are still interacting with the target nucleus and each other. 
The basic idea was to solve the time-dependent Schr\"odinger equation over the finite period of time when the pulse was acting on the target and to then effectively propagate the solution to infinite time by using the propagated wavepacket as the source term in a time-independent driven Schr\"odinger equation with the field-free Hamiltonian. The method proposed in I was illustrated with computation of one-and two-photon ionization cross sections for atomic hydrogen and we outlined the theory for extending the method to two-electron targets. 

In this paper, we demonstrate the viability of the approach with computations on atomic helium. We will show that  the present method allows us to extract fully differential ionization probabilities over the entire bandwidth of the pulse and to resolve structures arising from relatively long-lived autoionizing states, which might require prohibitively long propagation times with traditional time-dependent approaches.  Although the methodology can be applied with arbitrary field strengths, we confine our attention here to low intensity fields so that we can compare the present method with the results of other studies that calculated one- and two-photon ionization cross sections in the perturbative limit. 

The outline of this paper is as follows. The theory is outlined in Sec.~\ref{Theory and Computation}, beginning with a derivation of the driven equation and the extraction of ionization amplitudes, followed by  explicit formulas for one- and two-photon cross sections for single and double ionization and a brief description of the computational procedures we employ. In Sec.~\ref{Single ionization of helium by one or two photons} we present results for one- and two-photon single ionization of helium, while Sec.~\ref{One photon double ionization} presents results for one-photon double ionization. We conclude with a brief discussion.
\section{Theory and Computation}
\label{Theory and Computation}
\subsection{Time-dependent Schr\"odinger equation and extraction of ionization amplitudes}
The methodology we use is fully detailed in I \cite{palacios07} and so only the essentials are repeated here. We assume the atom, initially in its ground state, is subjected to a time-varying pulse that starts at $t=0$ and ends at $t=t_\textrm{final}$. To track the time evolution of the wave function during this period, we solve  the time-dependent Schr\"odinger equation:
\begin{equation}
i\frac{\partial}{\partial t}\Psi(t) = \EuScript{H}(t)\Psi(t)\,.
\label{eq:TDSE}
\end{equation} 
At $t= t_\textrm{final}$, the time-varying field ends  and the wave function continues to evolve under the time-independent atomic  Hamiltonian; this time evolution can be written explicitly as:
\begin{equation}
\Psi(t) = e^{-iH(t-t_\textrm{final})}\Psi(t_\textrm{final})\quad t>t_\textrm{final}\,.
\label{eq:TDwfcn0}
\end{equation}
We next define a scattered wave $\Psi_{sc}$ by taking the Fourier transform, from $t_\textrm{final}$ to infinity, of Eq.~(\ref{eq:TDwfcn0})
\begin{equation}
\begin{split}
\Psi_{sc}&\equiv -ie^{-iEt_\textrm{final}}\int_{t{_\textrm{final}}}^{\infty}dt e^{i(E+i\epsilon)t}\Psi(t)\\
&=-i\int_{0}^{\infty}dt e^{i(E+i\epsilon -H)t}\Psi(t_\textrm{final})\\
&=\frac{1}{(E+i\epsilon-H)}\Psi(t_\textrm{final})\\
&=G^+\Psi(t_\textrm{final})\,,
\end{split}
\end{equation} 
or, equivalently,
\begin{equation}
\label{eq:DSE}
(E-H)\Psi_{sc}=\Psi(t_\textrm{final})\,.
\end{equation}
Thus the scattered wave, from which we will extract physical information, satisfies a driven Schr\"odinger equation in which the propagated wavepacket at the end of the pulse appears as the source term.

To construct amplitudes for single and double ionization, we begin by formally expanding the wavepacket, at the end of the pulse, in the complete set of bound, single and double continuum  eigenstates of the target Hamiltonian:
\begin{equation}
\begin{split}
\Psi &(\mathbf{r}_1,\mathbf{r}_2,t_\textrm{final})=\\
&\psi_{\rm{bound}}(\mathbf{r}_1,\mathbf{r}_2)+\psi_{\rm{single}}(\mathbf{r}_1,\mathbf{r}_2)+\psi_{\rm{double}}(\mathbf{r}_1,\mathbf{r}_2)\\
&=\psi_{\rm{bound}}(\mathbf{r}_1,\mathbf{r}_2)\\
&+\sum_n\int dk^3_n  \, C(\mathbf{k}_n) \psi^{-}_{\mathbf{k}_n}(\mathbf{r}_1,\mathbf{r}_2)\\
&+\int dk^3_1 \int dk^3_2 \, C(\mathbf{k}_1,\mathbf{k}_2) \psi^{-}_{\mathbf{k}_1,\mathbf{k}_2}(\mathbf{r}_1,\mathbf{r}_2)  \, ,
\end{split}
\label{eq:2eexpansion}
\end{equation}
where $\psi_{\rm{bound}}(\mathbf{r}_1,\mathbf{r}_2)$  contains the contributions from the bound states of the target, $n$ runs over the bound states of He$^+$ and the coefficients $C(\mathbf{k}_n)$ and $C(\mathbf{k}_1,\mathbf{k}_2)$ are amplitudes for single and double ionization, respectively.
As we showed in I, substituting Eq.~(\ref{eq:2eexpansion}) into the expression $\Psi_{sc}=G^+\Psi(t_\textrm{final})$, and using the asymptotic form of the full Green's function, allows us to write the asymptotic forms of $\Psi_{sc}$ for single \cite{Hostler} and double ionization \cite{Kadyrov03a}:
\begin{equation}
\begin{split}
\Psi_\textrm{sc}^{\textrm{\scriptsize{single}}}&\underset{r_1 \rightarrow \infty} \sim \sqrt{2 \pi} \sum_n C(k_n\hat{\mathbf{r}}_1) \frac{e^{i(kr_1 + (Z/k) \ln 2kr_1)}}{r_1}\phi_n(\mathbf{r}_2) \\
\Psi_\textrm{sc}^{\textrm{\scriptsize{double}}}&\underset{\rho \rightarrow \infty} \sim
\sqrt{2 \pi i}\left( \frac{K^3}{\rho^5} \right)^{1/2} C(k_1\hat{\mathbf{r}}_1,k_2\hat{\mathbf{r}}_2)
e^{iK \rho +\zeta\ln 2 K \rho}
\, ,
\end{split}
\label{eq:asymp}
\end{equation}
where
\begin{equation}
\zeta =  \frac{Z}{k_1}+\frac{Z}{k_2} -\frac{1}{|\mathbf{k}_1-\mathbf{k}_2|}\,,
\end{equation}
$\phi_n$ is a bound state of He$^+$ and $\rho=\sqrt{r_1^2+r_2^2}$.

By solving the driven equation (\ref{eq:DSE}) using exterior complex scaling (ECS), pure outgoing boundary conditions are automatically imposed on the scattered wave function. And having identified the amplitudes $C(\mathbf{k}_n)$ and $C(\mathbf{k}_1,\mathbf{k}_2)$ in the asymptotic form of the scattered wave, their explicit evaluation is done in terms of the following surface integrals \cite{TopicalReview2004}:
\begin{equation}
\begin{split}
C(& \mathbf{k}_n) =  \frac{1}{2} \int\big[\phi^{-*}_\mathbf{k}(\mathbf{r}_1) \phi_n^*(\mathbf{r}_2)( \nabla \Psi_\textrm{sc}(\mathbf{r}_1,\mathbf{r}_2))\\
&- \Psi_\textrm{sc}(\mathbf{r}_1,\mathbf{r}_2)
\nabla 
 (\phi^{-*}_\mathbf{k}(\mathbf{r}_1) \phi_n^*(\mathbf{r}_2)
\big]\cdot dS\, ,
 \end{split}
 \label{eq:1esurface}
\end{equation}
for single ionization, and
\begin{equation}
\begin{split}
C(\mathbf{k}_1,&\mathbf{k}_2) 
= \frac{1}{2} e^{i\chi} \int  \left[\phi^{-*}_\mathbf {k_1}(\mathbf{r_1}) \phi^{-*}_\mathbf {k_2}  \, \nabla \,  \Psi_\textrm{sc} (\mathbf{r}_1,\mathbf{r}_2) \right. \\
& \left. \qquad \qquad  -\Psi_\textrm{sc} (\mathbf{r}_1,\mathbf{r}_2)\, \nabla  \,(\phi^{-*}_\mathbf {k_1}(\mathbf{r_1}) \phi^{-*}_\mathbf {k_2}(\mathbf{r_2}))  \,  \right]  \cdot d\mathbf{S} \, ,
\end{split}
\label{eq:3Damp}
\end{equation} 
for double ionization, where $\nabla=(\nabla_1,\nabla_2)$ and $\chi$ is a volume-dependent phase that makes no contribution to any physical observable~\cite{palacios07}. The testing functions $\phi^-_{\mathbf{k}}$ are momentum-normalized Coulomb functions with a nuclear charge $Z=2$.

\subsection {One-photon cross sections }
The amplitudes for ionization extracted via Eq.~(\ref{eq:1esurface}) or  Eq.~(\ref{eq:3Damp}) will generally depend on the parameters (intensity, bandwidth, etc.) of the radiation pulse that produced the wavepacket being analyzed. However, if the intensities are such that time-dependent perturbation theory gives an accurate description of the physical process, then the amplitudes can be used to construct  one-photon cross sections and, if the pulse durations are not too short, two-photon cross sections,  over the range of energies within the bandwidth of the pulse.

In the dipole approximation, the laser-atom interaction in the velocity gauge is given in terms of the electron's momentum operator $\mathbf{p}$ and the vector potential $\mathbf{A}$  by $U(\mathbf{r},t)=(e/mc){\bf A}(t)\cdot{\bf p}$. For a photon energy $\omega$ and a total pulse duration $T$, $\mathbf{A}(t)$ may be written
\begin{equation}
\mathbf{A}(t)= \left\{
\begin{array}{cl} A_0 F_{\omega}(t) \boldsymbol{\epsilon}\, & t\in [0,T] \\ 0\, & {\rm elsewhere}
\end{array}
\right.
\label{eq:vectorpotential}
\end{equation}
where $ \boldsymbol{\epsilon}$ is the polarization vector.  We choose a sine squared envelope for the time and frequency dependence of the pulse, $F_\omega(t)$, 
 \begin{equation}
F_\omega(t) = \sin^2\left(\frac{\pi}{T}t\right) \sin(\omega t)\, .
\label{eq:Fwt}
\end{equation}

One-photon ionization cross sections are obtained using first-order time-dependent perturbation theory (TDPT). In first-order TDPT, the transition amplitude, $C^{1\omega}$, between an initial state of energy $E_i$ and a continuum final state of energy $E_f$, caused by a pulse of duration $T$ characterized as in Eqs.(\ref{eq:vectorpotential}) and (\ref{eq:Fwt})  is
\begin{equation}
C^{1\omega} =\frac{-i e A_0}{\hbar m c}  \langle \Phi^-_f|{\bf \epsilon}\cdot {\bf p}|\Phi_i\rangle 
\widetilde{F}^{1\omega}(\omega,\Delta E,T) 
\label{CKuno}
\end{equation}
where $\Delta E=(E_f-E_i)$ and
\begin{equation}
\begin{split}
\widetilde{F}^{1\omega}(\omega,&\Delta E,T) = \int_{0}^T  e^{i\Delta Et/ \hbar}  F_\omega(t)\\
&=\frac{e^{-i \omega T}
\left(e^{i (\omega -\Delta E/\hbar)T} -1 \right) \pi^2}{(T^2(\omega-\Delta E/\hbar)^2 - 4\pi^2)(\omega-\Delta E/\hbar)}
\end{split}
\end{equation}
and
$\langle \Phi^-_{f}|{\bf \epsilon}\cdot {\bf p}|\Phi_i\rangle$ is the dipole matrix element connecting the initial and final states. 

The cross section for one-photon single ionization, leaving the ion in state $n$ is
\begin{equation}
\frac{d\sigma^{1\omega}}{d\Omega} = \frac{4\pi^2 \alpha k_n }{m \Delta E}
| \langle \Phi^{-}_\mathbf{k_n}|\boldsymbol{\epsilon}\cdot \mathbf{p}|\Phi_i \rangle|^2\, ,
\label{sigma-one-PT-vel}
\end{equation}
while the one-photon double ionization cross section is
\begin{equation}
\frac{d\sigma^{1\omega}}{dE_1 d\Omega_1d\Omega_2} = \frac{4\pi^2 \alpha k_1 k_2 }{\hbar^2 \Delta E}
| \langle \Phi^{-}_{\mathbf{k_1},\mathbf{k_2}}|\boldsymbol{\epsilon}\cdot \mathbf{p}|\Phi_i \rangle|^2\, .
\label{sigma-double-PT-vel}
\end{equation}
Combining Eqs.~(\ref{sigma-one-PT-vel}) and (\ref{sigma-double-PT-vel}) with Eq.~(\ref{CKuno}) then gives 
\begin{equation}
\frac{d \sigma^{1\omega}}{d \Omega} = \frac{ 4 \pi^2 k_n e^2 m}{\alpha \Delta E }
\frac{|C(\mathbf{k_n})|^2}{|A_0|^2 |\widetilde{F}^{1\omega}(\omega,\Delta E,T)|^2 } \, ,
\label{sigma-one-vel}
\end{equation}
and 
\begin{equation}
\frac{d\sigma^{1\omega}}{dE_1 d\Omega_1d\Omega_2} = \frac{ 4 \pi^2 k_1 k_2 e^2 m^2}{\alpha \hbar^2 \Delta E }
\frac{|C(\mathbf{k_1},\mathbf{k_2})|^2}{|A_0|^2 |\widetilde{F}^{1\omega}(\omega,\Delta E,T)|^2 } \, .
\label{sigma-double-vel}
\end{equation}
Thus the factorability of the  transition probability  in first-order TDPT  allows us to extract the one-photon differential single and double ionization cross sections from a single pulse within its bandwidth, defined as the range of photon frequencies for which  $\widetilde{F}^{1\omega}(\omega,\Delta E,T)$  is appreciably  nonzero. 

It is worth pointing out that for long pulse durations $|\widetilde{F}^{1\omega}(\omega,\Delta E, T)|^2$ tends to $(3\pi T/16)\delta (\omega-\Delta E/\hbar)$ \cite{palacios07,Piraux2006}. This identity may be used  to define cross sections at $\omega=\Delta E/\hbar$ via Fermi's golden rule, but this is only meaningful when the long time limit has been reached. Equations~(\ref{sigma-one-vel}) and (\ref{sigma-double-vel}), on the other hand, properly define the cross sections for any length pulse - a point that has evidently been missed in the recent literature~\cite{pindzola2007}.

\subsection {Two-photon cross sections }

For a two-photon absorption process, we can use TDPT in second order to write the amplitude for a transition 
between an initial state of energy $E_i$ and a continuum final state of energy $E_f$ as
\begin{equation}
\begin{split}
 C^{2\omega} = 
  \left(\frac{-i\alpha A_0}{e m }\right)^2
&\sum_m   \langle \Phi^-_f|  \boldsymbol{\epsilon} \cdot \mathbf{p} | \Phi_m \rangle  \langle \Phi_m | \boldsymbol{\epsilon} \cdot \mathbf{p} | \Phi_i \rangle\\
&\times F^{2\omega}(E_f,E_m,E_i,\omega,T)\, ,
\end{split}
\label{eq:CKdos}
\end{equation}
where the sum $m$ is over all the eigenstates of the target. The 
coefficient $F^{2\omega}(E_f,E_m,E_i,\omega,T)$ is given by
\begin{equation}
\begin{split}
F^{2\omega}(E_f,&E_m,E_i,\omega,T)=\nonumber\\
&\frac{1}{2} \int_0^T dt'
e^{i (\Delta E_{fm}/\hbar -\omega) t'}  \sin^2(t' \pi /T) \nonumber \\
&\times  \frac{1}{2} \int_0^{t'} dt'' 
e^{i (\Delta E_{mi}/\hbar  -\omega)t''}  \sin^2(t'' \pi /T) \,
\end{split}
\end{equation}
where $\Delta E_{ij}=E_i-E_j$.\\

To connect Eq.~(\ref{eq:CKdos}) with the familiar expression for the two-photon cross section, we define a reduced coefficient or "shape function":
\begin{equation}
\begin{split}
\widetilde{F}^{2\omega}(E_f,&E_m,E_i,\omega,T)=\\
&(E_i+\Delta E_{fi}/2-E_m)
F^{2\omega}(E_f,E_m,E_i,\omega,T).
\end{split}
\label{eq:reduced}
\end{equation} 
As we explained in I, we have found that if  the photon frequency $\omega$ is not too close to being in resonance with a transition to one of the intermediate states, $m$, then the shape function $\widetilde{F}^{2\omega}$ is well
approximated by an expression that does not depend on the energies of the intermediate states in the sum in Eq.~(\ref{eq:CKdos}) and which becomes exact in the long $T$ limit:
\begin{widetext}
\begin{equation}
\label{eq:longT}
\begin{split}
\widetilde{F}^{2\omega}(E_f,E_m,E_i,&\omega,T)  \approx
 \widetilde{\mathfrak{F}}(E_f,E_i,\omega,T)\\
&=\frac{6 e^{-i T (2 \omega -\Delta E_{fi})} \left(-1+e^{i T (2 \omega -\Delta E_{fi})}\right) \pi ^4}{(2 \omega -\Delta E_{fi} ) \left[T^4 (2 \omega -\Delta E_{fi})^4-20 \pi ^2 T^2 (2 \omega -\Delta E_{fi} )^2+64 \pi ^4\right]}\, ,
\end{split}
\end{equation}
\end{widetext}
giving
\begin{equation}
\begin{split}
 C^{2\omega} \approx
  \left(\frac{-i\alpha A_0}{e m }\right)^2
& \langle \Phi^-_f|  \boldsymbol{\epsilon} \cdot \mathbf{p}\frac{1}{(E_i+\Delta E_{fi}/2-H)}\boldsymbol{\epsilon} \cdot \mathbf{p} | \Phi_i \rangle\\
&\times \widetilde{\mathfrak{F}}(E_f,E_i,\omega,T)\, ,
\end{split}
\label{eq:CKapprox}
\end{equation}

We can now connect the amplitudes we extract from the wavepacket using Eqs.~(\ref{eq:1esurface}) or (\ref{eq:3Damp}) with  differential cross sections  for two-photon ionization.  The two-photon single ionization cross section is given by the expression,
\begin{equation}
\begin{split}
\frac{d \sigma^{2\omega}}{d \Omega} &= \frac{(2 \pi)^3  k_n \alpha^2\hbar^3}{(\Delta E_{fi}/2)^2 m^3}\\
&\times \left| 
\langle \Phi_\mathbf{k_n}^-|   
{\bf \epsilon}\cdot {\bf p}
(E_i +\Delta E_{fi}/2-H)^{-1}
{\bf \epsilon}\cdot {\bf p}
|\Phi_i\rangle
\right|^2 \, ,
\end{split}
\end{equation}
while the two-photon double ionization cross section is given by 
\begin{equation}
\begin{split}
&\frac{d \sigma^{2\omega}}{dE_1 d \Omega_1d \Omega_2} = \frac{(2 \pi)^3  k_1 k_2 \hbar \alpha^2}{(\Delta E_{fi}/2)^2 m^2}\\
&\times \left| 
\langle \Phi_\mathbf{k_n}^-|   
{\bf \epsilon}\cdot {\bf p}
(E_i +\Delta E_{fi}/2-H)^{-1}
{\bf \epsilon}\cdot {\bf p}
|\Phi_i\rangle
\right|^2 \, .
\end{split}
\end{equation}
Combining the previous two cross section  definitions with Eq.~(\ref{eq:CKapprox}) then gives
\begin{equation}
\frac{d \sigma^{2\omega}}{d \Omega} =
 \frac{(2 \pi)^3 \hbar^3 k_n \alpha^2}{(\Delta E_{fi}/2)^2 m^3}
\frac{ | C(\mathbf{k_n})|^2}
{ \left(\frac{e^2 |A_0|^2}{m^2 c^2 \hbar^2}\right)^2 |\widetilde{\mathfrak{F}}(E_f,E_i,\omega,T)|^2 } \, ,
\label{eq:2hvxsec}
\end{equation}
and 
\begin{equation}
\frac{d \sigma^{2\omega}}{dE_1d \Omega_1d \Omega_2} =
 \frac{(2 \pi)^3 \hbar k_1 k_2 \alpha^2}{(\Delta E_{fi}/2)^2 m^2}
\frac{ | C(\mathbf{k_1},\mathbf{k_2})|^2}
{ \left(\frac{e^2 |A_0|^2}{m^2 c^2 \hbar^2}\right)^2 |\widetilde{\mathfrak{F}}(E_f,E_i,\omega,T)|^2 } \, .
\label{eq:2hv2e}
\end{equation} 
Once again, the factorability of the transition probability allows us to extract cross sections from a single pulse within its bandwidth, but we must emphasize that, in contrast to the one-photon case, this factorability is approximate and, as we shall see,  breaks down when the pulse width is very short or may require very long propagation times to resolve the energy dependence of the cross section when the photon frequency is close to being in resonance with a discrete intermediate state.

\subsection{Implementation}
We employ the same computational techniques here that we have used our recent work on two-electron problems~\cite{Vanroose06, Yip2007}. The two-electron wave function is first expanded in products of spherical harmonics
\be
\begin{split}
\label{eqn:partialwave}
 \Psi(\mathbf{r}_1,\mathbf{r}_2,t)=\sum_{l_1m_1}\sum_{l_2m_2}&
\frac{1}{r_1r_2}\psi_{l_1m_1,l_2m_2}(r_1,r_2,t)\\
&\times Y_{l_1m_1} (\hat{\mathbf{r}}_1)Y_{l_2m_2}(\hat{\mathbf{r}}_2)\, .
\end{split}
\ee
We include all $lm$-pair configurations that can be constructed using some specified value of $l_{max}$. Integration over the angular variables then gives a set of coupled equations for the two-dimensional radial functions $\psi_{l_1m_1,l_2m_2}(r_1,r_2,t)$. The radial degrees of freedom are in turn discretized using a finite-element, discrete variable representation (FEM-DVR) with a product basis of Lobatto shape-functions~\cite{dvr00}. The value of $l_{max}$, as well as the size and density of grid points required for convergence, will depend on the photon energy as well as the process under consideration. 

Exterior complex scaling of the radial coordinates,
\be
\label{eqn:ecs}
r \rightarrow
\begin{cases}
r, & r \le R_0 \\
R_0+(r-R_0)e^{i\theta}, &  r>R_0,
\end{cases}
\ee
defines a radius $R_0$ beyond which the radial coordinates are complex-valued. The round-state wave function $\Psi_0$, which serves as the initial wavepacket for the time-propagation, is obtained by diagonalizing the field-free Hamiltonian on a relatively small portion of the real grid ($r_{max}\approx 30$ Bohr), with configurations representing total angular momentum $L=0$. This wavepacket is then propagated  on the {\em real part} of the grid over the duration of the pulse. The time-propagation is carried out using a Cranck-Nicholson propagator,
\begin{equation}
\label{eq:CN}
(1-\frac{i \Delta t}{2}H) \Psi(\mathbf{r}_1,\mathbf{r}_2,t+\Delta t)=(1+\frac{i \Delta t}{2}H) \Psi(\mathbf{r}_1,\mathbf{r}_2,t)
\end{equation}
with a time step $\Delta t$ on the order of $10^{-3}$ atomic units. Since the Hamiltonian is time-dependent, Eq.~(\ref{eq:CN}) requires inverting a large matrix at each time step. However, since the wavepacket changes little with each $\Delta t$, we have found that an iterative solution at each step, which requires only matrix-vector multiplications, converges in several iterations. Since the time-propagation is carried out on the real part of the grid, $R_0$ must be chosen large enough to contain the spreading wavepacket over the duration of the pulse and avoid unphysical reflections from the grid boundaries. The value of $R_0$ required will generally increase with photon energy; in these calculations, we found that $R_0=130$ Bohr was sufficient for the range of photon energies considered.

\begin{figure}[h]
\begin{center}
\resizebox{!}{0.95\columnwidth}{\includegraphics*{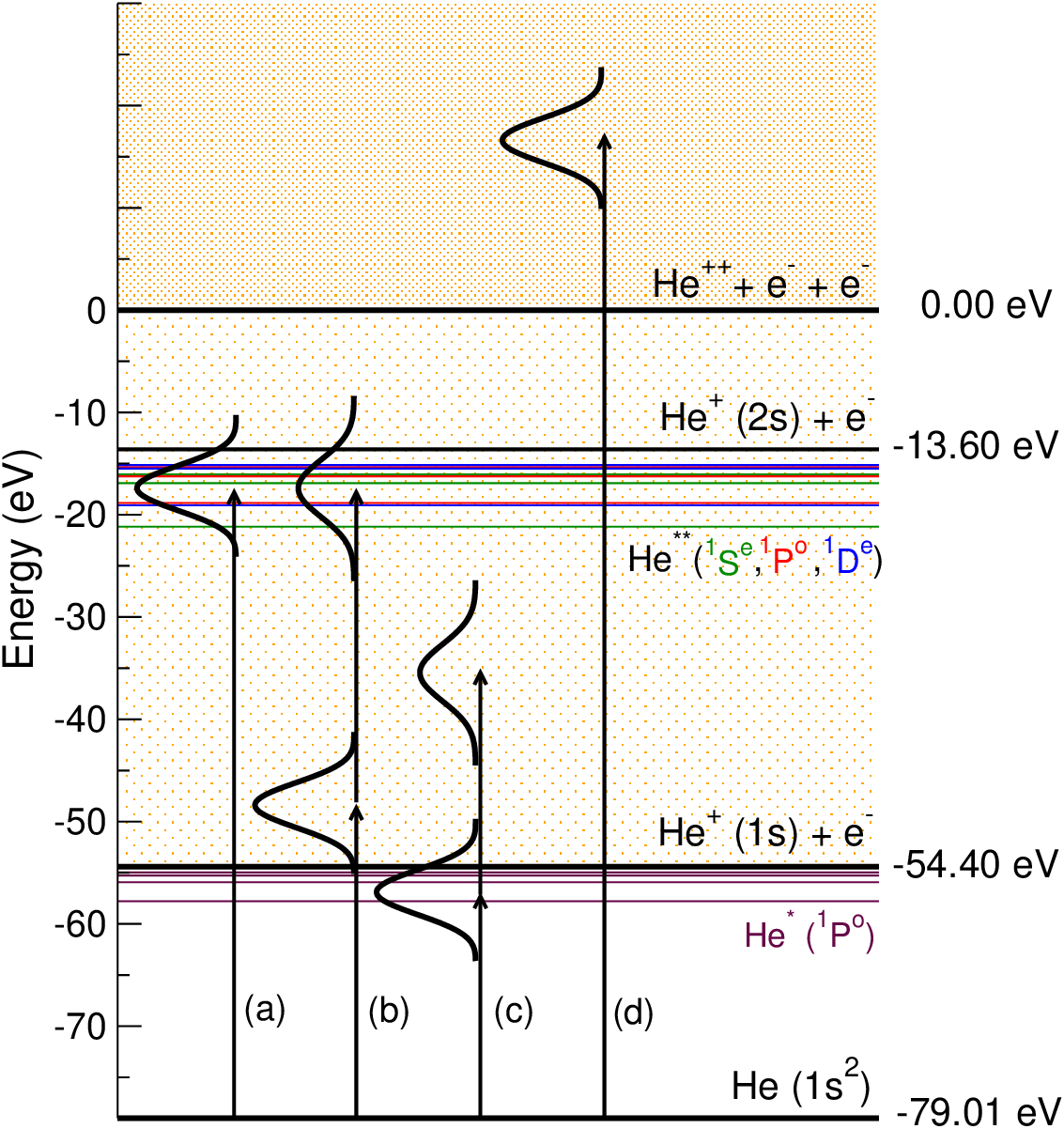}}
\caption{(Color online) Energetics of helium ionization. Schematic representation of (a) one-photon single ionization process, (b) and (c) two-photon single ionization processes; and (d) one-photon double ionization process.} 
\label{graph1}
\end{center}
\end{figure}
The wavepacket at the end of the pulse serves as the driving term for the scattered wave equation (Eq.~(\ref{eq:DSE})). This equation is solved on the full exterior scaled grid and provides the scattered wave from which the physical amplitudes are extracted, as outlined above. We reiterate that the time propagation is carried out only once for a particular laser pulse and then the scattered wave equation can be solved for any energy within the pulse bandwidth.
\section{Single ionization of helium by one or two photons}
\label{Single ionization of helium by one or two photons}
We restrict our calculations to weak fields where perturbation theory can be applied and, therefore, a cross section defined. In this way, we can check the accuracy of the method by comparing with existing calculations and experimental measurements. 

All results reported here were obtained with an intensity I$=10^{12}$ W cm$^{-2}$, which is high enough to provide relatively large ionization rates for one and two photon transitions and low enough to keep the processes within the perturbative regime \cite{palacios06,palacios07}.

Firstly, we look at  single ionization  by absorption of one or two photons using pulses of duration T=$0.9$ fs. Over the range of photon energies considered, calculations are converged by including pairs of spherical harmonics built with individual angular momenta up to $l_{max}=3$ and leading to total angular momenta up to $L=2$. In order to be consistent, we construct our ground state with the same maximum value for individual angular momenta. In the weak field limit, optical selection rules apply, so only channels with total symmetry $^1$S$^{\textrm{e}}$, $^1$P$^{\textrm{o}}$ and $^1$D$^{\textrm{e}}$ ($L=0$, 1 and 2) are accessible. 
 \begin{figure}
\begin{center}
\resizebox{!}{1.15\columnwidth}{\includegraphics*{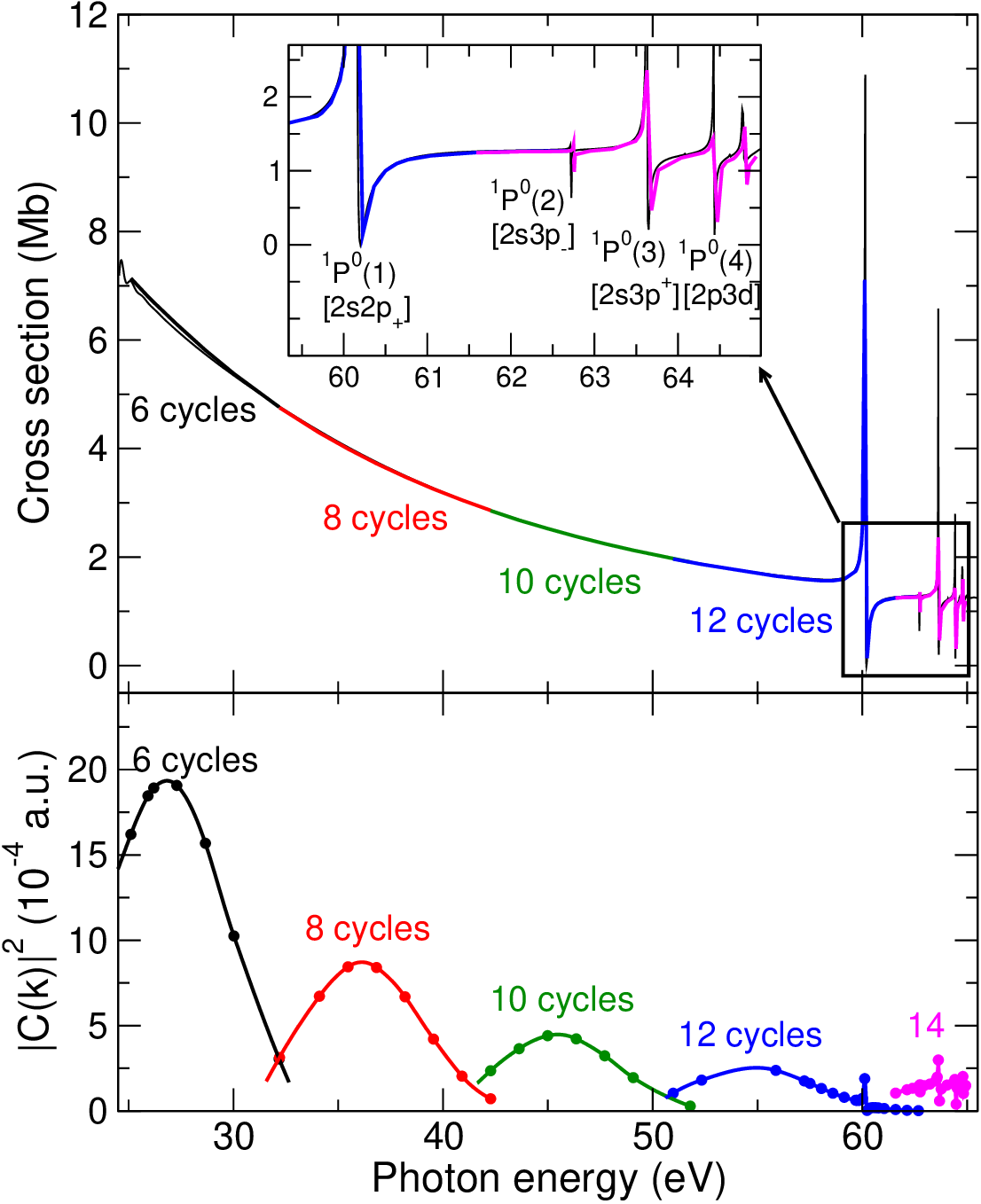}}
\caption[]{(Color online) Top panel: One-photon single ionization cross sections, in units of megabarns, as a function of photon energy. 1Mb=10$^{-18}$cm$^2$. Thin black full line: time independent perturbation theory results from reference \cite{McCurdyMartin04}. Thick lines: Present results for T$=0.9$ fs, $I=10^{12}$ W cm$^2$ and different number of cycles. Inset in top panel: Enlargement in the region where doubly excited states are resolved. Bottom panel: Squared amplitudes, in atomic units, from which the total cross sections in the top panel were extracted.} 
\label{graph2}
\end{center}
\end{figure}
\subsection{Resolving autoionization resonances in one-photon single ionization}
Figure ~\ref{graph1} shows a schematic representation of one- and two-photon ionization processes in atomic helium. One-photon ionization, labeled (a), is possible with photon energies above 24.6 eV. For two-photon single ionization, with a threshold at half this energy, we can distinguish above-threshold ionization (ATI), labeled (b), in which the first absorbed photon is above 24.6 eV, from case (c), where two photons are required to ionize the target. For photon energies above 79.01 eV, one photon double ionization, labeled (d), is possible.

The total cross section for one-photon single ionization is plotted in the top panel Fig.~\ref{graph2} as a function of photon energy. These cross sections were extracted from calculations using different central frequencies of the field, i.e. different numbers of optical cycles. For a pulse duration T$=0.9$ fs, 6 cycles correspond to a pulse of central frequency around $27.6$ eV, 8 cycles correspond to $36.76$ eV, and so on. For the cross sections depicted in Fig.~ \ref{graph2}, the corresponding squared amplitudes, appearing in the bottom panel, capture the energy bandwidth of the pulse, i.e., the Fourier transform of $F_\omega(t)$ defined in Eq.(\ref{eq:Fwt}). 

The results plotted in  Fig.~\ref{graph2} are practically indistinguishable from time independent perturbation theory calculations (black thin line) and experimental measurements given in ref.~\cite{McCurdyMartin04}. Agreement is excellent even in the region between 60 and 65 eV where one-photon absorption can populate doubly-excited states of $^1P^0$ symmetry, the first four of which have been labeled  in the inset of the top panel. Cross sections in this region were obtained from calculations using pulses of $0.9$ fs with 12 and 14 optical cycles. 

We should point out that conventional time-dependent treatments would require long propagation times  to fully resolve the autoionizing structures seen in Fig.~\ref{graph2} -- greater than 6ps, for example, in the case of the $^1P^o$ resonance near 62.7 eV. With the present method, we can obtain such results with much shorter propagation times because the time span from the end of the pulse to infinity -- when the two electrons are still interacting -- is handled exactly by Eq.~\ref{eq:DSE}.

\subsection{Two photon single ionization}
\begin{figure}[h]
\begin{center}
\resizebox{!}{1.10\columnwidth}{\includegraphics*{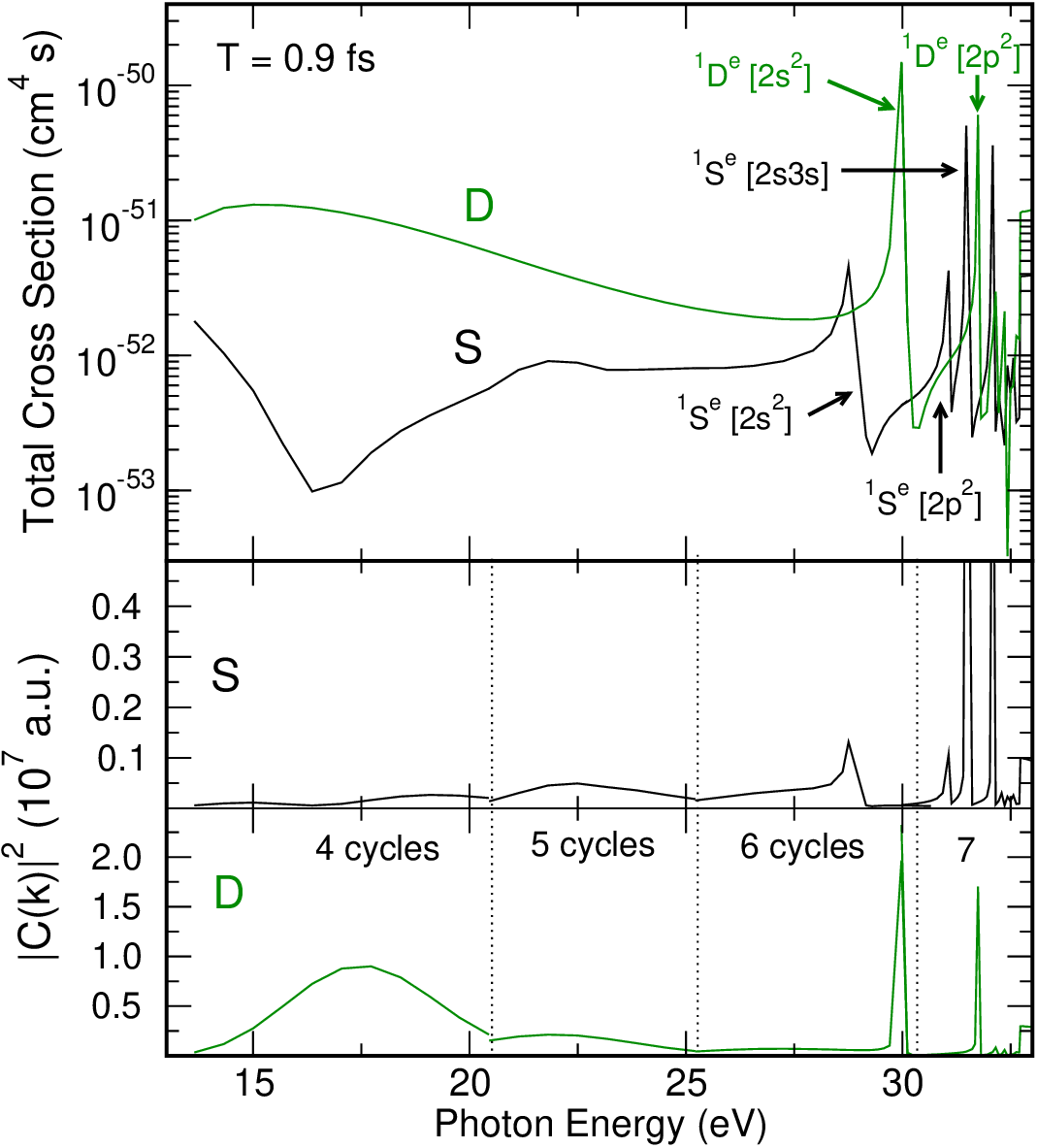}}

\caption[]{(Color online) Upper panel: S and D symmetry components of the total cross section for two-photon single ionization of helium, calculated with  pulses of duration of T=0.9 fs and $I=10^{12}$ W cm$^{-2}$ and varying number of optical cycles (4, 5, 6 and 7). Lower panel: Squared amplitudes, in atomic units, from which the total cross sections in the top panel were extracted.} 
\label{graph3}
\end{center}
\end{figure}

We have also explored single ionization of helium by two-photon absorption. Total cross sections for this process are plotted in Fig. \ref{graph3} as a function of photon energy. Given the optical selection rules, only states with $^1$S$^{\textrm{e}}$ and $^1$D$^{\textrm{e}}$ final symmetries will be populated from the ground state. The corresponding amplitudes appear in the botton panels for various pulses with different optical cycles.

Structures appearing in the total cross section for photon energies above 28 eV correspond to doubly excited states of symmetries $^1$S$^{\textrm{e}}$ and $^1$D$^{\textrm{e}}$ (labeled in the figure) which decay to the continuum after the pulse is turned off, as we have discussed in the previous subsection. We note that, at these photon energies, above threshold ionization (ATI) processes are taking place [see energy scheme of Fig.~\ref{graph1}, process labeled (b)]. The positions and  widths of these autoionizing states are in reasonable agreement results obtained from time-independent perturbation theory calculations in refs.  \cite{Lambropoulos2001,Hasbani2000,Sanchez95}. 

We must point out that our calculation with a 0.9 fs pulse does not reproduce any of the structures between 20 and 24.5 eV found with time-independent perturbation theory \cite{Lambropoulos2001}. At these energies, bound excited states of helium are populated by one-photon absorption [process represented in Fig.~\ref{graph1} as (c)], which leads to divergences in the total cross section when the pulse length goes to infinity. Since we are using a finite length pulse with an energy bandwidth wider than the energy spacing between these intermediate bound states, these cannot be resolved. 

\begin{figure}[h]
\begin{center}
\resizebox{!}{0.8\columnwidth}{\includegraphics*{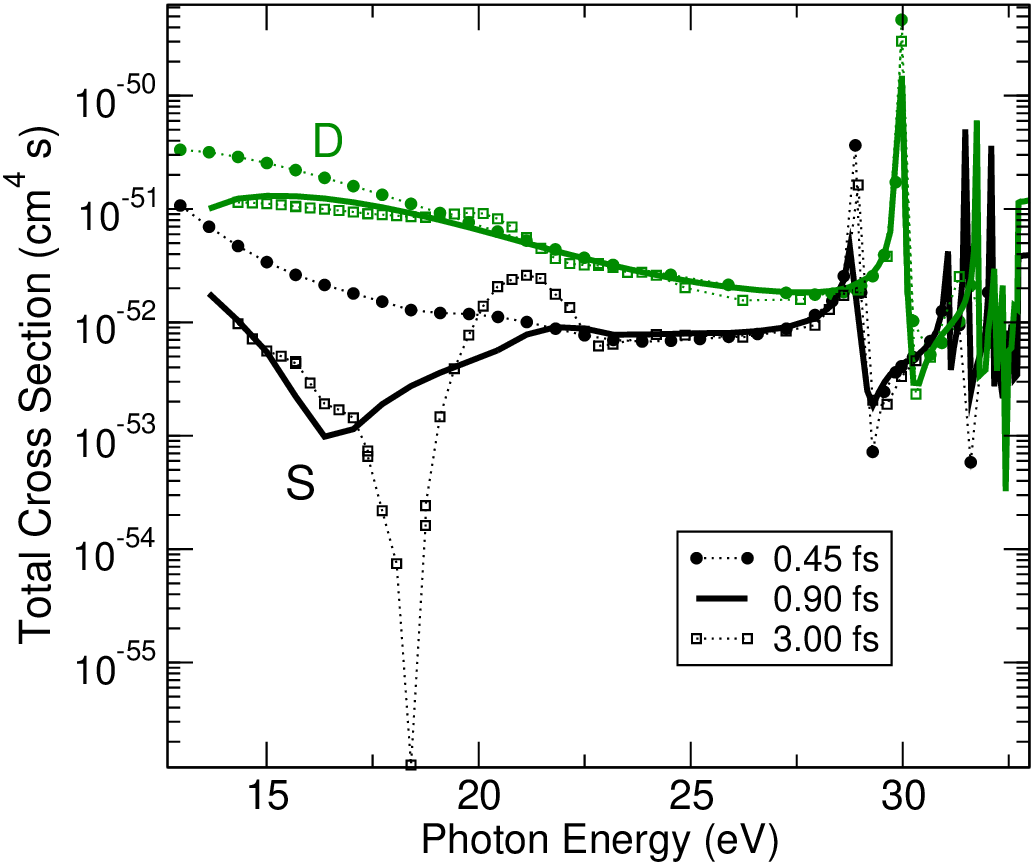}}
\caption[]{(Color online) Dependence of extracted two-photon single ionization cross section on pulse length.} 
\label{graph4}
\end{center}
\end{figure}

Figure~\ref{graph4} shows two-photon single ionization cross sections extracted from calculations using pulses of different durations. These results highlight the completely different nature of the structures arising from single-photon absorption by intermediate bound states and the structure associated with doubly excited states. The latter appear at photon energies above 25 eV where the cross sections are seen to be invariant to increasing the pulse length. The doubly excited states are first populated, in the presence of the field, by two-photon absorption  and  decay later in time. By contrast, the track of intermediate states between 19 eV and the first ionization threshold only begins to appear in the extracted cross sections for long pulse durations. With a short pulse duration of 0.45 fs, the effective bandwidth is $\Delta\omega = 2\pi/$T$ = 9.2$ eV, which is too broad to resolve any structure below 25 eV. For a 3 fs pulse, the bandwidth is $\sim 1.4$ eV and we see structure beginning to develop below 25 eV. Much longer pulse durations would be required to fully resolve these  structures. 

So to conclude this section, we reiterate that in the two-photon case, there are two different issues with respect the pulse duration that must be carefully considered. First, as we have just discussed, is the fact that with  finite pulses intermediate bound states will be resolved only if the energy bandwidth of the pulse is narrower than the structures in question. This is simply the physics of the problem. The second point is that the factorability of a "shape function"  from the transition probability for two-photon absorption relies on an approximation (Eq.~(\ref{eq:longT})) that becomes unreliable for very short pulses. Indeed, careful examination of Fig.~\ref{graph4} shows that in the case of a 0.45 fs pulse, there are small errors in the computed cross sections even above 25 eV. We hasten to remark that these issues are only relevant when one wishes to compute cross sections to compare with the results of time-independent perturbation theory. Whether or not we operate in the perturbative limit, the amplitudes computed for any pulse length or field intensity will still give correct ionization probabilities.

\section{One photon double ionization}
\label{One photon double ionization}
\begin{figure}[htbp]
\begin{center}
\resizebox{!}{0.9\columnwidth}{\includegraphics*{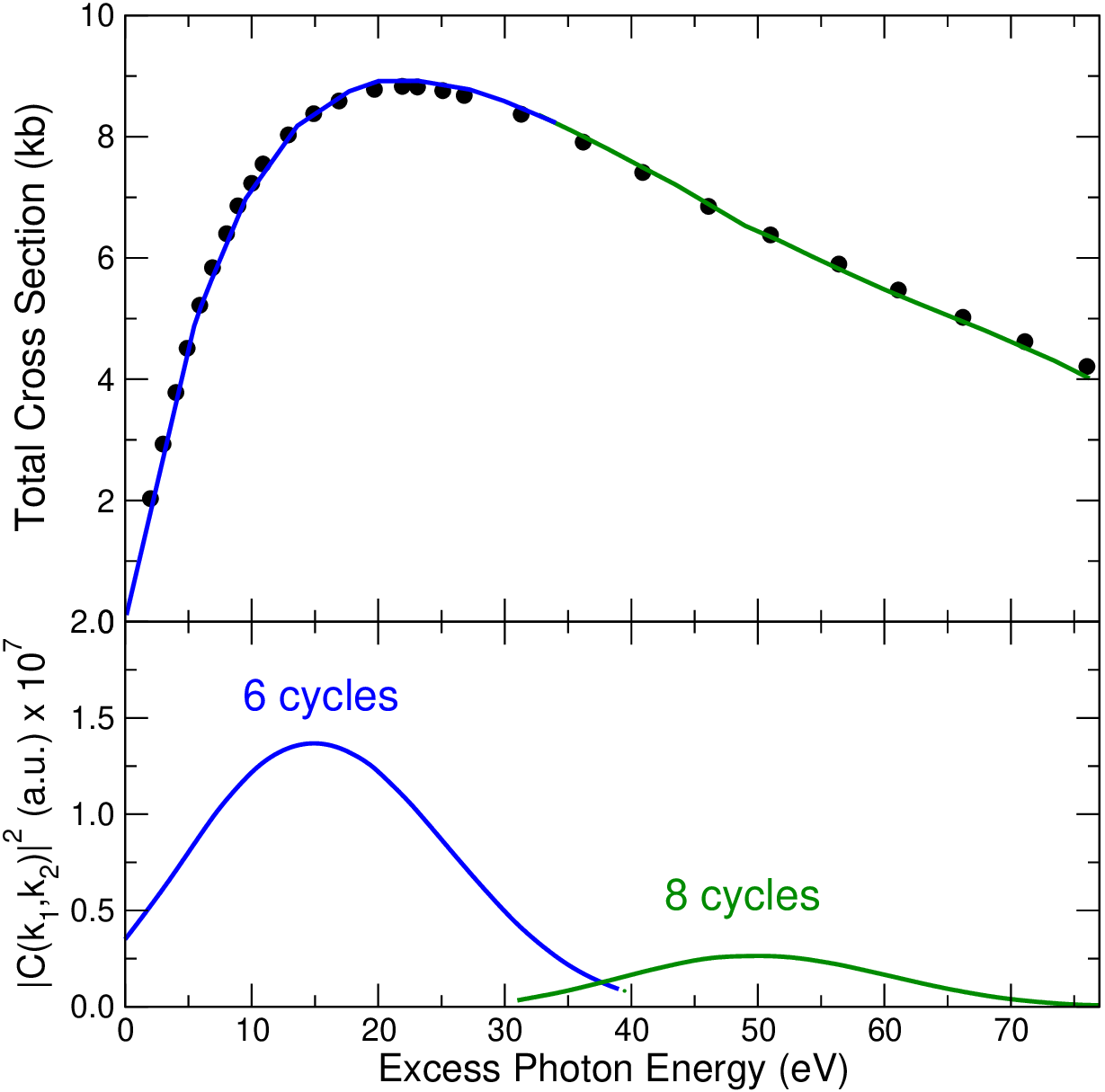}}
\caption[]{(Color online) Upper panel: total cross section, in units of kilobarns, for one-photon double ionization of helium. 1 kb=10$^{-21}$cm$^2$. Solid curve: present results; dots: experimental results of Samson~\cite{Samson98}.  Lower panel: squared amplitudes, in atomic units, from which the total cross sections in the top panel were extracted.}
\label{graph5}
\end{center}
\end{figure}

A key test of the method is the extraction of total and differential cross sections for double ionization processes. In Fig.~\ref{graph5} we show the total cross section for one-photon double ionization (process labeled  (d) in  Fig.~\ref{graph2}). The total cross section was calculated for a wide range of photon energies by using only two different wavepackets propagated with a 250 attosecond pulse. 
The agreement with experimental results by Samson~\cite{Samson98} is excellent. Once again, it is the exact factorability of the amplitude in the one-photon case that allows us obtain accurate results with such a short pulse duration.

\begin{figure}[htbp]
\begin{center}
\resizebox{!}{0.7\columnwidth}{\includegraphics*{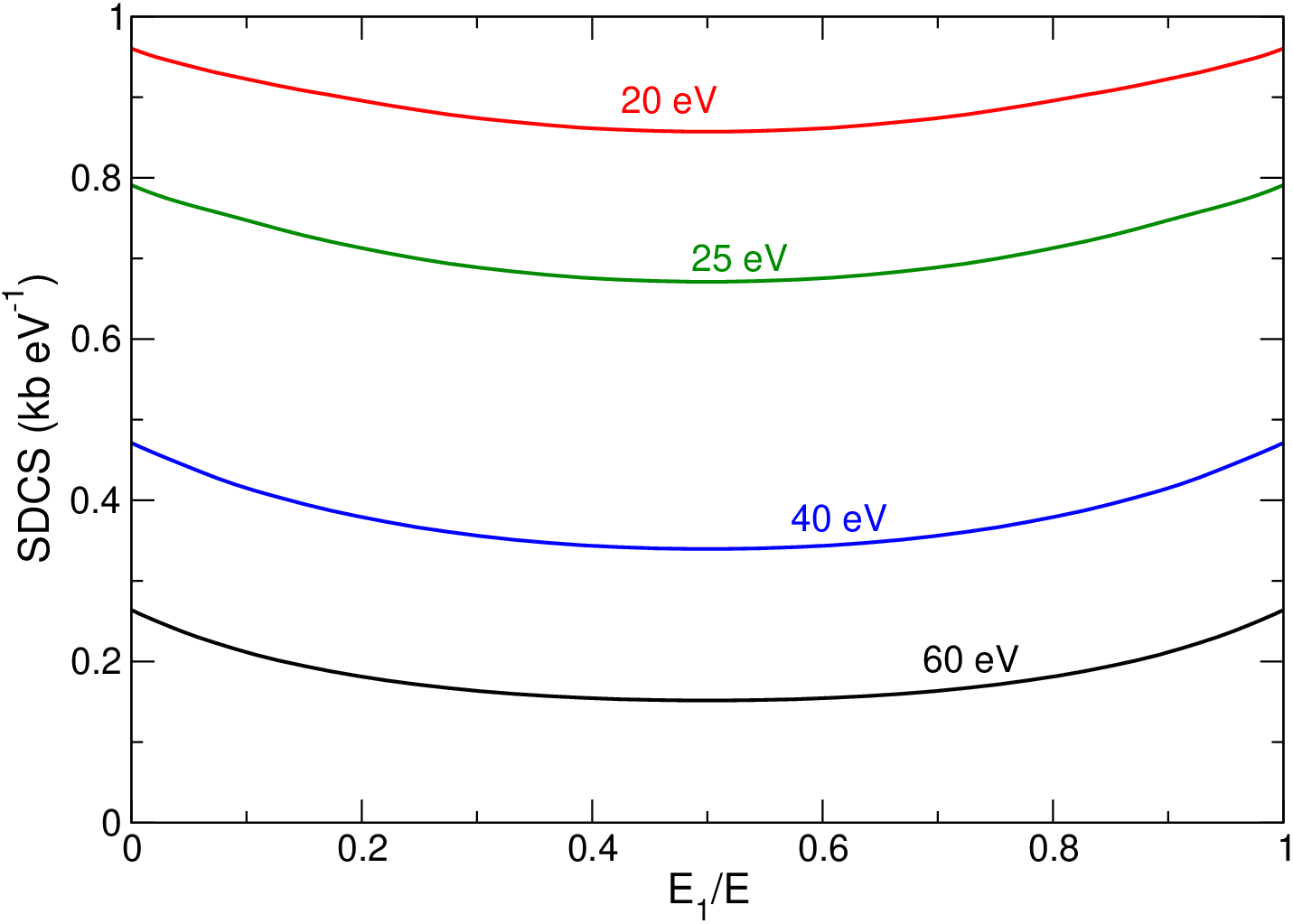}}
\caption[]{(Color online) Single differential cross sections, in units of  kilobarns per eV, for one-photon double ionization of helium at different total energies for the ejected electrons.}
\label{graph6}
\end{center}
\end{figure}

Since one-photon double ionization processes are extremely sensitive to electron correlation effects in both initial- and final-states , single and triple differential cross sections can be sensitive to higher values of electron orbital angular momentum. We have checked that, for the range of energies considered here, convergence is achieved by including $l_{max}=4$ for individual angular momenta in the spherical harmonic basis for the propagating wavepacket as well as for the ground state.

 In Fig.~\ref{graph6}, we plot the single differential cross section (SDCS), as a function of energy sharing,    for four different total energies. Our results  are in excellent agreement with previous experimental measurements \cite{wehlitz1991}. as well as the calculations of Colgan and Pindzola \cite{colgan02}.

\begin{figure}[htbp]
\begin{center}
\resizebox{!}{1.0\columnwidth}{\includegraphics*{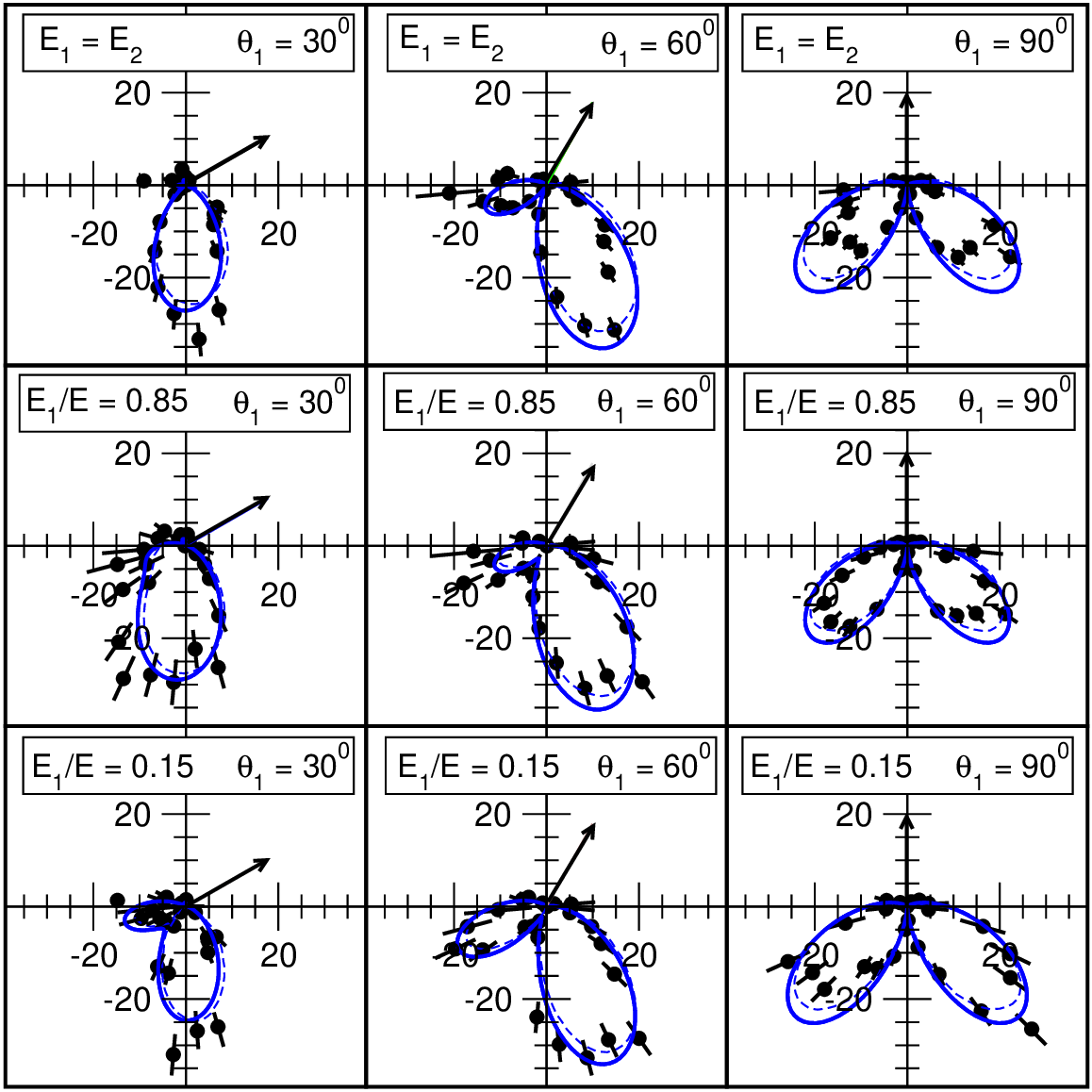}}
\caption[]{(Color online) Triple differential cross sections (TDCS), in units of  barns per eV per steradian$^2$, for one photon double ionization of helium at 20 eV above threshold.  1 barn=10$^{-24}$m$^2$.  Solid curves: present results; broken curves: CCC results of Kheifets and Bray~\cite{kheifetsBray98}; points: experimental results of Br\"auning {\em et al.}~\cite{Brauning98}.}
\label{graph7}
\end{center}
\end{figure}

In Fig.~\ref{graph7}, we show triple differential cross sections (TDCS) at 20 eV above the double ionization threshold, where the total cross section reaches its maximum value. 
Most of the experiments, including those shown here, have been performed in coplanar geometry, i.e., with the polarization vector and both momenta $\vec{k}_1$ and $\vec{k}_2$ in the same plane.
For the TDCS, the time propagation is carried out with a 6 cycles pulse (central frequency of 99.98 eV), and the extraction for a total absorbed energy of 99.01 eV. Results are shown for three different fixed values of ejection angle  for electron 1($30^\circ, 60^\circ$ and $90^\circ$) and three different energy sharings. We find excellent agreement with the absolute experimental results of Br\"auning {\em et al.}~\cite{Brauning98}, as well as the convergent close-coupling (CCC) calculations of Kheifets and Bray~\cite{kheifetsBray98}. 
 
\begin{figure}[htbp]
\begin{center}
\resizebox{!}{0.65\columnwidth}{\includegraphics*{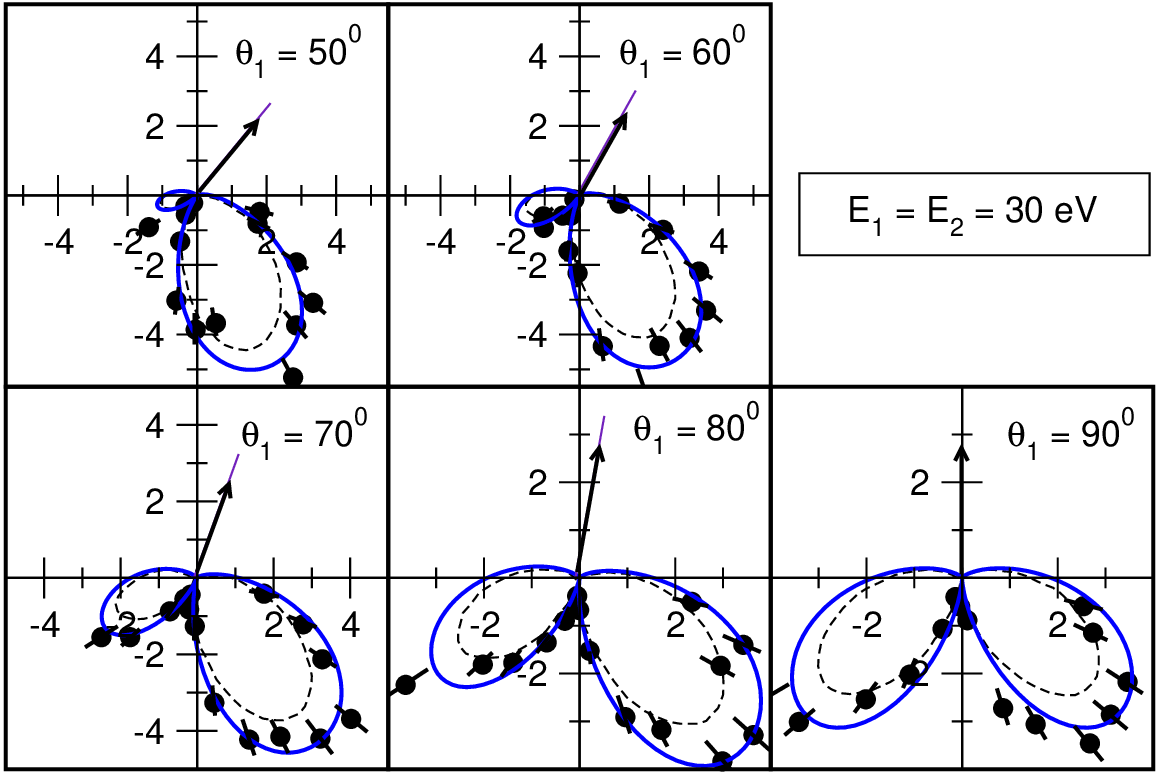}.}
\caption[]{(Color online) As in Fig.~\ref{graph7}, at 60 eV above threshold. Solid curves: present results. Dots: experimental results ~\cite{Dawson2001}, normalized to the present results. Broken curves: CCC  from 
ref.~\cite{Dawson2001}}
\label{graph8}
\end{center}
\end{figure}

Figure~\ref{graph8} shows TDCS results at 60 eV above threshold for equal energy sharing of the ejected electrons. The experimental results in this case~\cite{Dawson2001} were internormalized but not absolute, so we normalized them to our present results for comparison. The CCC theoretical results for this case are seen to agree with our calculations in shape, but are $\sim$20 percent smaller in magnitude. Time-dependent close coupling calculations ~\cite{colgan02} are indistinguishable from our results (for this reason these results are not plotted).

\begin{figure}[htbp]
\begin{center}
\resizebox{!}{1.0\columnwidth}{\includegraphics*{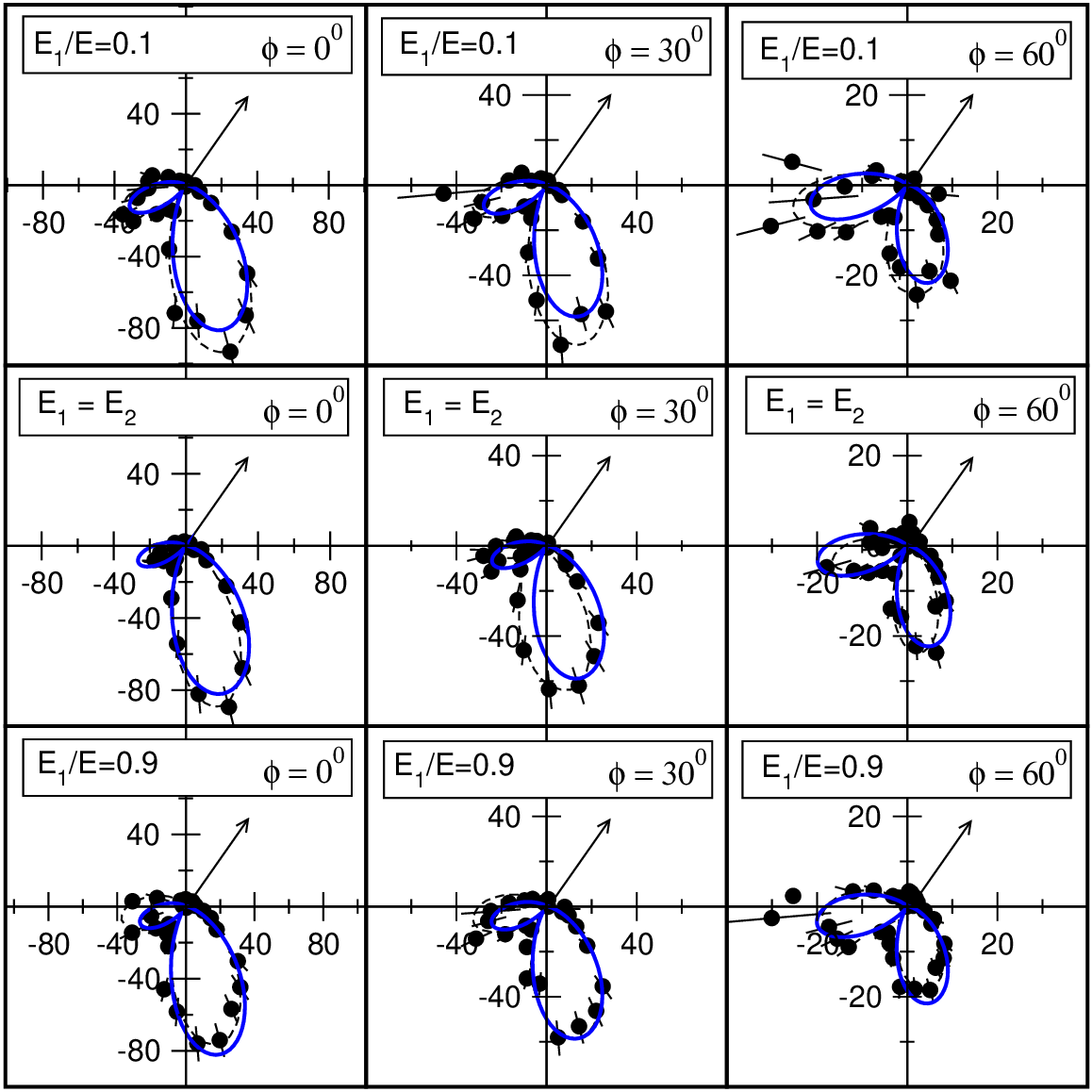}}
\caption{(Color online) As in Fig.~\ref{graph7}, at 6 eV above threshold. Solid curves: present results; experimental results (dots) and fourth-order Wannier theory results (broken curves) are from ref.~\cite{DornerPRA1998}. }
\label{graph9}
\end{center}
\end{figure}

Finally, we present TDCS results for a photon energy of 85 eV, which is only 6 eV above the ionization threshold. The TDCS are plotted in Fig.~\ref{graph9}, along with absolute experimental results of Do\"rner {\em et al.}~\cite{DornerPRA1998}. In the experiment, the data was binned over finite ranges of energy sharing( 0-0.2, 0.5 and 0.8-1) and polar angle $\theta_1$ (40-65$^\circ$) for the fixed electron. Our calculations are for  fixed energy sharings of 0.1, 0.5 and 0.9 and we fixed the polar angle $\theta_1$ at 45$^\circ$. In Fig.~\ref{graph9}, each row corresponds to a different energy sharing while the columns correspond to  a different value of azimuthal angle $\phi$ between the fixed and varied electrons. The calculations were done for $\phi=0^\circ, 30^\circ \,\, {\rm and}\,\, 60^\circ$, while in the experiment  $\phi$ was binned (0-20$^\circ$, 40-65$^\circ$ and 45-90$^\circ$). We find excellent agreement with experiment.  Also shown in Fig.~\ref{graph9} are the results from ref.~\cite{DornerPRA1998} obtained using fourth-order Wannier theory~ \cite{Feagin95}, which do not provide absolute cross sections but evidently do reproduce the shapes of the TDCS at these low energies.

\section{Discussion}
We have demonstrated that the method we proposed in I can indeed be applied to the study of single and double ionization processes in two-electron systems. By working with relatively low field intensities where perturbation theory is expected to be valid, we have been able to show that the cross sections we calculate for atomic helium are consistent  with earlier results obtained using time-independent methods, as well as experimental data. By propagating a wavepacket in the presence of a single pulse over its time duration, we can extract cross sections over the entire bandwidth of the pulse, even in the ATI region.

Having demonstrated that we can reproduce the results of time-independent calculations with this method, we hasten to add that our main purpose in developing this approach is to provide an efficient and reliable method for the  exploration of problems that are difficult or impossible to study with time-independent techniques. Such problems include studies of above-threshold, two-photon double ionization, which are far from straightforward with time-independent perturbative methods~\cite{Horner2007he2p}, short-pulse, intense field studies where perturbation theory is not valid, and simulations of two-color, pump-probe experiments which require an approach that is explicitly time-dependent.

%
%
%
%
\begin{acknowledgments}
This work was performed under the auspices of the US Department of Energy
by the University of California Lawrence Berkeley National Laboratory
under Contract DE-AC02-05CH11231 and
was supported by the U.S. DOE Office of Basic Energy
Sciences, Division of Chemical Sciences. CWM acknowledges support from the National Science Foundation (Grant No. PHY-0604628).
\end{acknowledgments}

%
%

\end{document}